\documentclass[a4paper,prb,10pt,notitlepage,twocolumn]{revtex4-1}

\usepackage{graphicx,color}
\usepackage[colorlinks=true,linkcolor=blue,citecolor=blue]{hyperref}
\usepackage{amssymb,amsmath,bbold,mathtools}

\begin{document}

\title{Probing the metallic energy spectrum beyond the Thouless energy scale using the singular value decomposition}
\author{Richard Berkovits}
\affiliation{Department of Physics, Jack and Pearl Resnick Institute, Bar-Ilan University, Ramat-Gan 52900, Israel}

\begin{abstract}
Disordered quantum systems feature an energy scale know as the Thouless energy.
For energy ranges below this scale, the properties of the energy spectrum can
be described by random matrix theory. Above this scale a different behavior
sets in. For  a metallic system  it has been long ago shown
by Altshuler and Shklovskii that the number
variance should increase as a power law with a power dependent only on the
dimensionality of the system. Although tantalizing hints for  this behavior
have been seen in previous numerical studies, it is quite difficult to verify
this prediction using the standard local unfolding methods. Here we use
a different unfolding method, i.e., the singular value decomposition, and
establish a connection between the power law behavior of the scree plot
(the singular values ranked by their amplitude) and the power law behavior
of the number variance. Thus we are able to numerically verify
the Altshuler and Shklovskii's prediction for disordered $3D$, $4D$, and $5D$
single-electron Anderson models on square lattices in the metallic regime.
The same method could be applied to systems such as the 
Sachdev-Ye-Kitaev model and various interacting many body models for which
the many body localization occurs. It has been recently reported that such
systems exhibit a Thouless energy and analyzing the spectrum's behavior
on larger scales is of much current  interest.
\end{abstract}


\maketitle

\section{Introduction}

Weakly disordered quantum systems are known to exhibit a universal
behavior of their energy spectrum which depends only on the
symmetry of the system \cite{mehta91}.
The statistical properties of the energy
spectrum do no depend on the details of the system, and are described by
a random matrix model with the same symmetry. This behavior is extremely
useful in identifying and understanding various properties of metallic
systems \cite{shklovskii93,ghur98,alhassid00,mirlin00,evers08}.

As pointed out by Altshuler and Shklovskii \cite{altshuler86},
this universal behavior
holds only for energy scales which are below the Thouless energy $E_T$.
The Thouless energy corresponds to $E_T=\hbar/t_T$ where the Thouless time
$t_T=L^2/D$ ($L$ is the linear dimension of  the sample, $D$ the diffusion
constant) depicts the time it takes for a  diffusing particle to sample all
the system. For shorter times the motion is not yet diffusive and therefore
dependent on details of the local system.
Thus, above this energy scale, the behavior of statistical properties
of the spectrum will diverge from the random matrix predictions.
The canonical measure used to probe this deviation is the number variance
\cite{mehta91}, defined as the variance in the number of energy levels
within an energy window for an unfolded energy spectrum.
Defining a window of size $E$, one can count
the number of levels within this window for a given realization of disorder
and obtain the average number of levels, $\langle n(E) \rangle$,
and the variance,
$\langle \delta^2 n(E) \rangle=\langle (n(E) - \langle n(E) \rangle)^2 \rangle$,
where $\langle \ldots \rangle$ denotes an average over an ensemble of
different realizations of disorder.
For the Wigner-Dyson random matrix ensemble (Gaussian orthogonal ensemble, GOE):
$\langle \delta^2 n(E) \rangle= 0.44+
(2/\pi^2)\ln(\langle n(E) \rangle)$, while for the localized regime
$\langle \delta^2 n(E) \rangle=\langle n(E) \rangle$.
Metallic systems correspond to the Wigner-Dyson random matrix 
predictions up to a energy window of size $E_T$ and
Altshuler and Shklovskii \cite{altshuler86} predict that for $E>E_T$
the number variance will follow
$\langle \delta^2 n(E) \rangle \propto \langle n(E) \rangle^{d/2}$,
where $d$ is the dimensionality.

The deviation of the number variance from  the random matrix logarithmic
behavior to a stronger than linear behavior at large energies has been
observed in metallic system \cite{braun95,cuevas97}.
Recently, it has gained much interest beyond
the traditional single-particle disordered systems. Stronger than linear
deviations of the number variance beyond a certain energy scale have
been seen in the context of 
Sachdev-Ye-Kitaev (SYK) model \cite{garcia16,garcia18},
many body localization systems \cite{bertrand16,sierant19,corps20,wang21}
and the generalized Rosenzweig-Porter
random matrix model \cite{rosenzweig60,kravtosov15}. In all these cases
it was argued that the energy for which the number variance becomes
stronger than linear corresponds to the inverse of the time scale for which
the motion can no more sample the whole phase space.

Although the prediction for the behavior of the number variance 
on scales larger than the Thouless energy is straight forward, it
is not easy to corroborate even for simple single-particle systems
such as the Anderson model with any degree  of certainty. We shall see
that the main problem is the local unfolding procedure, as has been
noted by previous studies \cite{relano02,sierant19}.

Here we intend to address this challenge of verifying
Altshuler and Shklovskii prediction \cite{altshuler86}.
We shall illustrate in detail that a straight forward
study of the number variance using local unfolding is fraught
with ambiguities. Therefore, it is clear that a different tack is needed.
Here we will suggest that a couple of new measures, which are based on
singular value decomposition (SVD) method. This method
has been used to classify whether a system follows 
Wigner or  Poisson statistics 
\cite{fossion13,torres17,torres18},
and recently to identifying non-ergodic extended signature in the
Rosenzweig-Porter model \cite{berkovits20}. 
As we shall demonstrate, using the SVD method to replace the short
range unfolding provides a clearer way to study the 
the behavior of the  energy spectrum beyond the Thouless energy.

The paper is organized as follows. In the next section (Sec. \ref{s1})
we define
the single particle Anderson model on a square lattice for different
dimensionalities. The following section (Sec. \ref{s2})
presents the numerical results
for the number variance using local unfolding
and discusses the challenges it presents.
Section \ref{s3} shows the use of the SVD  to tease out the behavior
of the large energy scales. In its first subsection (\ref{s3a}) we give
an overview of the SVD method. In the next subsection
(\ref{s3b}) we use the SVD to perform a global  unfolding by filtering out
the  low modes of the singular values which corresponds to the global
features of  the  energy spectrum
and thus retaining only the fluctuations. Establishing analytically a connection
between the Power spectrum of these fluctuations  and the number variance
enables us to glean the long range behavior of the energy spectrum.
Since there is a connection  between the power spectrum and the
scree plot of the singular value modes (the singular values ranked by their
amplitude), as established in subsection \ref{s3b}, the scree plot may be used
to read off the long range spectrum properties. This is used in subsection
\ref{s3c} in order  to verify the dependence of the number variance
on dimensionality. Issues relating to the number of eigenvalues taken into
account and the number of realizations of disorder considered are also
discussed. In section \ref{s4} we discuss the possibility of applying the
SVD to additional interesting  systems such as
Sachdev-Ye-Kitaev model and disordered interacting many body models known to
exhibit many body localization. 

\section{Model}
\label{s1}

We consider a simple one-particle Anderson model on a $d$ dimensional
square lattice with sites at $\vec r = j_x\hat{x}+j_y \hat{y} + \ldots$,
where $j_i = 1,2,\ldots L_i$, and $L_i$  is the length in the $\hat i$
direction. Each site has
an on-site energy $\epsilon_{\vec r}$ chosen randomly from a box
distribution in the range $-W/2 \ldots W/2$. Nearest neighbor hopping between
the sites is considered, with a hopping matrix  element set to one. Thus the
Hamiltonian is written as:
\begin{eqnarray} \label{hamiltonian} 
  H = \sum_{\vec r} \epsilon_{\vec r} c_{\vec r}^{\dag} c_{\vec r} +
  \sum_{\vec r} \sum_{\hat a} c_{\vec r + \hat a}^{\dag} c_{\vec r},
\end{eqnarray}
where $c_{\vec r}^{\dag}$ is the creation operator at site $\vec r$ and
$\hat a = \pm \hat x,  \pm \hat y \ldots$ are unit vectors to the nearest
neighbor sites.

This model is known to exhibit a metal-insulator transition at
a critical disorder $W_C=16.5$ for the 3D case, $W_C=34.5$
for the 4D case, $W_C=57.5$ for the 5D case \cite{tarquini16}. In order
to study the long range spectra behavior deep in the metallic regime, we
concentrate on values of disorder much lower than the critical disorder,
i.e., $W=5$ and  $W=10$, .
Using exact diagonalization we calculate the eigenvalues for the
$L^d \times L^d$ matrices, where we consider hyper-cubes of size
$L=L_x=L_y,\ldots$  and hard wall boundary conditions.
For the 3D case we consider sizes $L=20,24,28$, corresponding to
$L^3=8000,13824,21952$, while
in the 4D case we evaluate sizes $L=9,10,11,12,13$, resulting in
$L^4=6561,10000,14641,20736,29561$,and
for the 5D case $L=6,7,8$, which amounts to
$L^5=7776,16807,32768$. Unless noted differently, in all cases the spectra
was calculated for $3000$ different realizations.

\section{Number Variance}
\label{s2}

To begin, we shall investigate the behavior of number variance at large
energies as function of the dimensionality and system size.
Using the $L^d$ eigenvalues, $\epsilon_i$,  obtained for each realization,
the spectrum is locally unfolded. The following local  unfolding was applied:
Each eigenvalue obtains the value
$\varepsilon_i=\varepsilon_{i-1}+2p(\epsilon_i-\epsilon_{i-1})/\langle \epsilon_{i+p}-\epsilon_{i-p}\rangle$
where $\langle \ldots \rangle$ is an average over realizations, and we have
checked that the results are not very sensitive to the value of $p$  (for
all results presented here $p=6$ was chosen).
The number variance is also averaged over $41$ positions of the
center of the energy window, $E(k)$,
equally spaced around the band center, where the
furthest point is no more than $1/15$ of the bandwidth from the center.
For each $E(k)$, the number of states in a window of width $E$ centered
at $E(k)$, 
$n_k(E)$ is evaluated, then the averages $\langle n(E) \rangle$
and $\langle n^2(E) \rangle$ are taken over all positions of the center $k$
and all realizations.

\begin{figure}
\includegraphics[width=8cm,height=!]{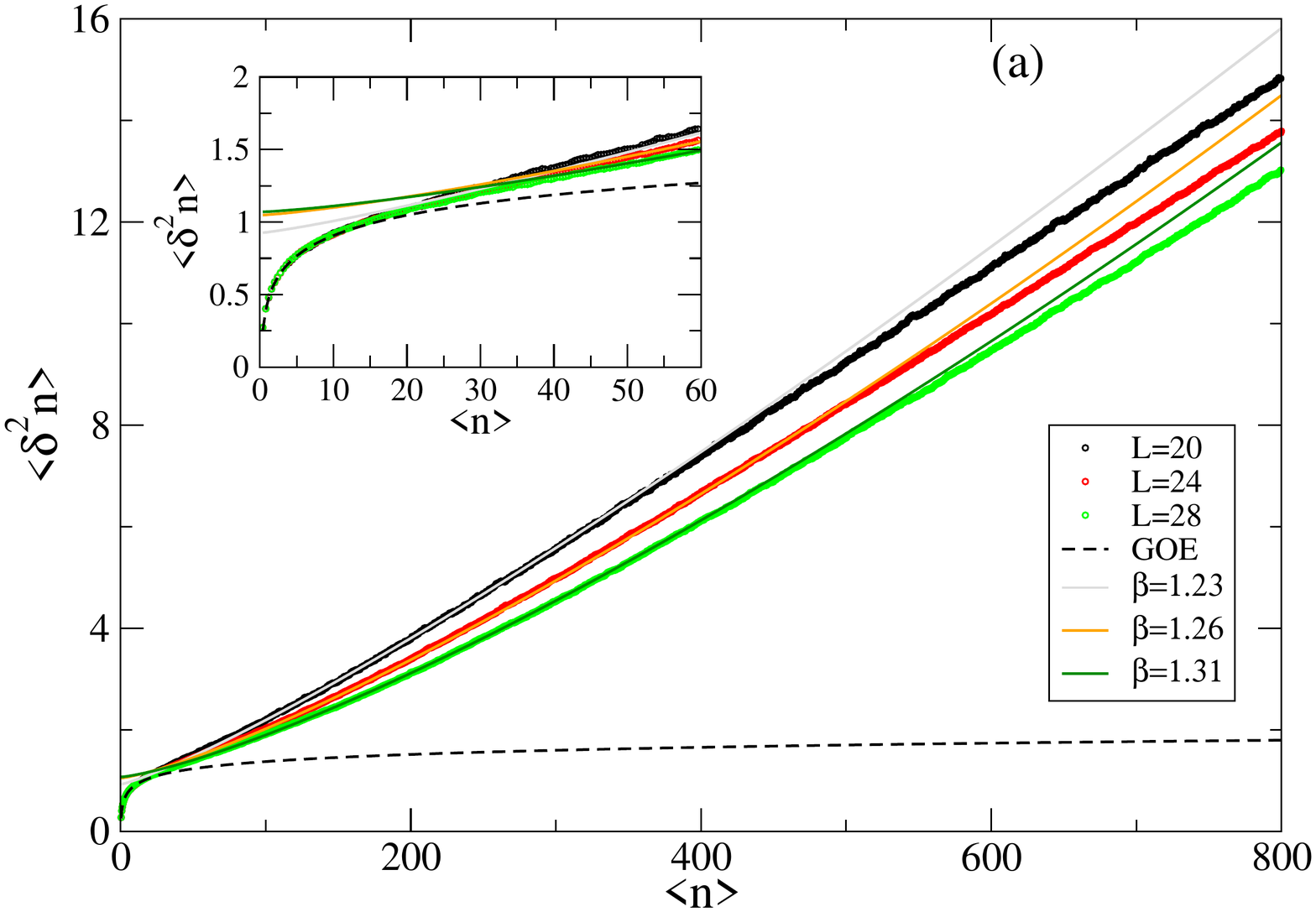}
\includegraphics[width=8cm,height=!]{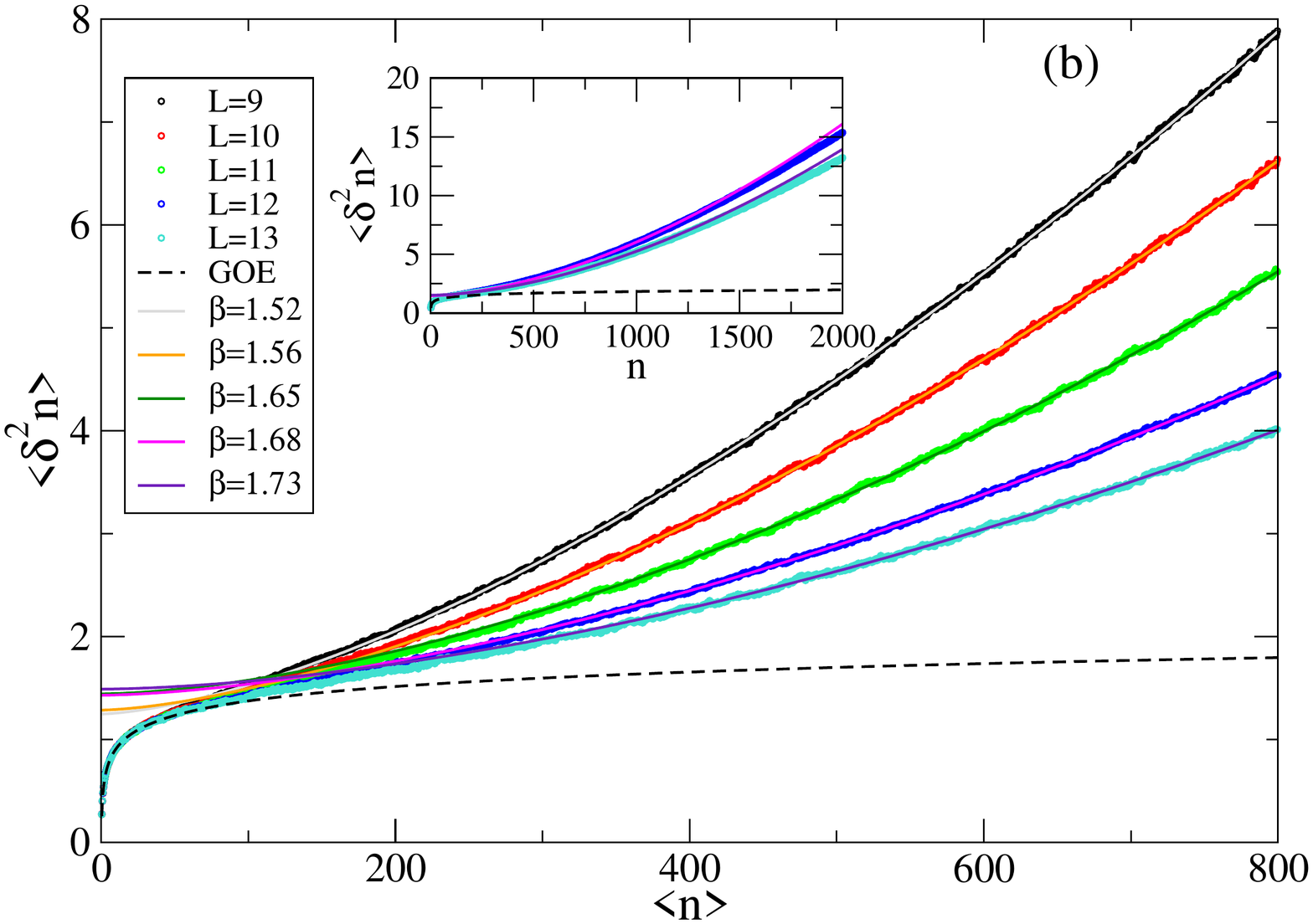}
\includegraphics[width=8cm,height=!]{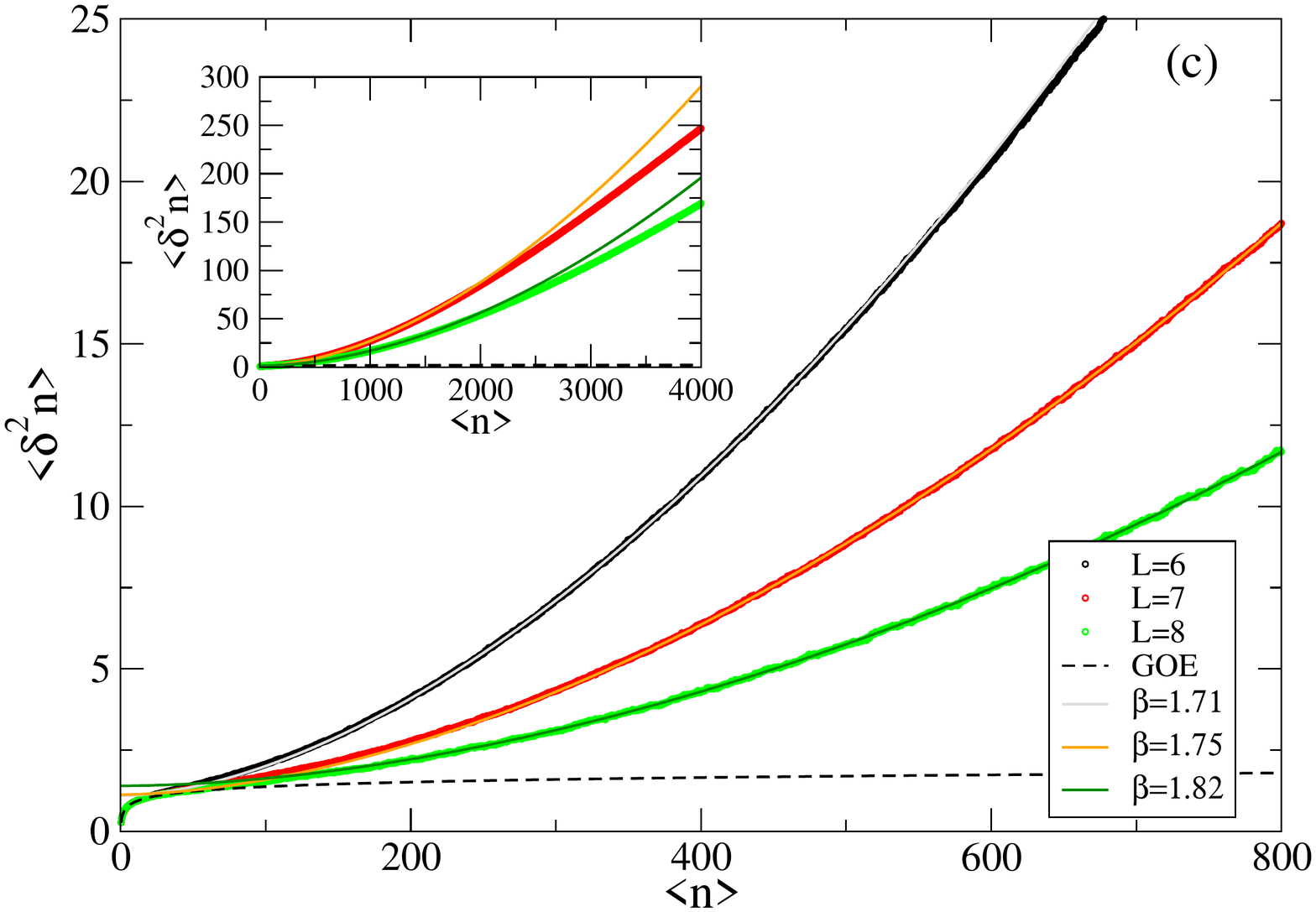}
\caption{\label{fig1}
  The variance $\langle \delta^2 n(E) \rangle$ as function of
  $\langle n(E) \rangle$ for 3D (a); 4D (b); and 5D (c) samples of
  different sizes. 
  (a) 3D samples
  of sizes $L=20,24,28$ and disorder $W=5$. For 
  $\langle n \rangle<\langle n(E_T) \rangle \sim 20$,
  Wigner Dyson (GOE) behavior is followed as can be seen clearly in the inset.
  In the region $50 <\langle n(E) \rangle < 150$ a fit to
  $\langle n(E) \rangle \sim \langle n(E) \rangle^\beta$
  is performed, resulting in $\beta=1.23, 1.26, 1.31$ for $L=20, 24, 28$
  correspondingly. Above $\langle n(E) \rangle > 80$, deviations from
  the power law become apparent and the variance increases
  more moderately. In the inset the deviation from the GOE logarithmic
  behavior at $\langle n(E_T) \rangle \sim 20$ can  be  clearly seen.
  (b) 4D samples of sizes $L=9, 10, 11, 12, 13$ and disorder $W=5$. For 
  $\langle n \rangle < \langle n(E_T) \rangle \sim 80$
  Wigner Dyson (GOE) behavior is followed.
  In the region $400 <\langle n(E) \rangle < 800$ a fit to
  $\langle n(E) \rangle \sim \langle n(E) \rangle^\beta$
  is performed, resulting in $\beta=1.52, 1.56, 1.65, 1.68, 1.73$ for
  $L=9, 10, 11, 12, 13$
  correspondingly. In the inset the behavior for larger values of the
  energy window is depicted.
  Clearly,  the variance does not continue to grow at the same pace.
    (c) 5D samples of sizes $L=7, 8$ and disorder $W=10$. For 
  $\langle n \rangle<\langle n(E_T) \rangle \sim 60$,
  Wigner Dyson (GOE) behavior is followed as can be seen clearly in the inset.
  In the region $200 <\langle n(E) \rangle < 400$ for $L=6$ and
  the region $400 <\langle n(E) \rangle < 800$ for $L=7,8$ a fit to
  $\langle n(E) \rangle \sim \langle n(E) \rangle^\beta$
  is performed, resulting in $\beta=1.75, 1.83$ for $L=7, 8$
  correspondingly. Again,  as shown in the inset,  for larger
  values of $\langle n(E) \rangle$ a weaker growth in variance appears.
 }
\end{figure}

One expects that for $E<E_T$, the number variance will follow the Wigner
Dyson
prediction, while for $E>E_T$, $\langle n^2(E) \rangle \sim
\langle n(E) \rangle^{d/2}$. This is probed in Fig. \ref{fig1} where
the variance $\langle \delta^2 n(E) \rangle$ as function of
$\langle n(E) \rangle$ for 3D, 4D, and 5D samples of different sizes are
plotted. In all cases the GOE logarithmic behavior is followed for
small energy windows. For energy windows larger than the Thouless
energy,
$\langle n \rangle>\langle n(E_T) \rangle$, a stronger than linear growth
sets in. The Thouless energy depends on disorder and dimensionality, and
we chose the  strength of disorder for each case
($W=5$ for 3D and 4D samples, $W=10$ for 5D  samples)
so $E_T$ will be such
that $\langle n(E_T) \rangle$, which corresponds to the dimensionless
conductance $g$, will be of order O(10-100). Above $\langle n(E_T) \rangle$ the
variance crosses over to a different behavior and shows a stronger than
linear increase.  Fitting the variance to a $\langle n(E) \rangle^\beta$
behavior shows that beyond the crossover region, there is a wide range for
which the power law $\beta$ is constant, and at even higher energies
deviations appear.

This general behavior is seen for all dimensionalities and system sizes.
For the 3D case with
$W=5$, the Thouless energy, i.e., where the variance starts to
diverges from the Wigner Dyson (GOE) predictions, appears around
$\langle n(E_T) \rangle \sim 20$. Fitting $\beta$ after the
variance has substantially diverged from the logarithmic behavior, i.e.,
in the region
$50 <\langle n(E) \rangle < 150$ results in
$\beta=1.23,1.26,1.31$ for $L=20,24,28$, which seems to hold well up to
$\langle n(E) \rangle < 350,400,450$ correspondingly. Above these values,
the numerical computed variance tapers off to a more moderate increase.
This  may be finite size effects or a problem with local unfolding
on larger energy scales.
For 3D we expect $\beta=1.5$, while
the values we see are below, but increasing
with the system size $L$. Thus, it may be that for much larger system
sizes the predicted value would be reached, nevertheless, extrapolating from the
change in $\beta$ as $L$ increases one concludes that much larger systems
will be needed in order to reach $\beta=1.5$.

A similar behavior is seen for higher dimensionality. For
the 4D case with the same disorder $W=5$ the Thouless energy is
larger and $\langle n(E_T) \rangle \sim 80$.
This is expected since as the number of nearest neighbors to which
the particle can hop increases the effect of disorder should decrease.
Again, we fit $\beta$ for the region for which the variance begins to
significantly diverge from GOE, $400 <\langle n(E) \rangle < 800$,
$\langle n(E) \rangle \sim \langle n(E) \rangle^\beta$,
and obtain $\beta=1.52,1.56,1.65,1.68,1.73$ for $L=9,10,11,12,13$.
Once more, values which are below the expected power law $\beta=2$, but
becoming closer as the system size is increase.
The same pattern emerges also for 5D samples where the disorder was increased
to  $W=10$ in order that the Thouless  energy will be similar to the
value obtained for lower dimensions $\langle n(E_T) \rangle \sim 60$
A fit for the region $400 <\langle n(E) \rangle < 800$ yields
$\beta=1.75,1.83$ for $L=7,8$, far from  the expected $\beta=2.5$
Once more, for 4D and 5D samples, for large energy scales,
$\langle n(E) \rangle > 1500$, the increase in the variance tappers.

Thus, although with much effort probing even larger sizes may be possible, it
nevertheless does not seem  very promising, and we shall turn in a different
direction. This direction is based on the singular value decomposition  method
and would be described in the following section.

\section{Singular Value Decomposition}
\label{s3}

\subsection{General}
\label{s3a}

Singular value
decomposition (SVD) is a mathematical method applied mainly in the field
of data analysis and has enjoyed growing popularity \cite{svd}.
In this method a matrix
$X$ of size $M \times P$
(not necessarily Hermitian nor square)
is decomposed to  a multiplication of
three matrices.
In general the relevant data is arranged by rows and columns,
where the specifics
depend on the application. Thus, $X$ is decomposed to $X=U \Sigma V^T$, here
$U$ and $V$  are $M\times M$ and $P \times P$ matrices,
while $\Sigma$ is a $M \times P$ {\it diagonal} matrix of rank
$r=\min(M,P)$. $\sigma_k$ stands for the $r$ diagonal elements of $\Sigma$ are
called the singular values (SV) of $X$. The SV are always positive
and could  be arranged by size so
$\sigma_1 \geq \sigma_2 \geq \ldots \sigma_r$.
The Hilbert-Schmidt norm of the matrix
$||X||_{HS}=\sqrt{Tr X^{\dag}X}=\sqrt{\sum_k \lambda_k}$
(where $\lambda_k=\sigma_k^2$).
Thus, $X$  can be written as a sum  of matrices $X^{(k)}$,
where $X_{ij}=\sum_k \sigma_k X^{(k)}_{ij}$, and
$X^{(k)}_{ij}=U_{ik}V^T_{kj}$.
Since the SV are ordered by amplitude, the main contribution to $X$ comes from
the first $m$ modes, and $X$ may be approximated by,
$\tilde X =\sum_{k=1}^m \sigma_k X^{(k)}$, for which $||X||_{HS}-||\tilde X||_{HS}$
is minimal. Thus, if $\lambda_k$ become relatively small for some value $m$
$\tilde X$ could be used as an approximation of $X$
\cite{svd1,svd2}.
Moreover, by plotting $\lambda_k$ as function of its ranking $k$
(known as a scree plot  in the context of statistical factor analysis
\cite{svd3}) one may gain some  insight into the statistical properties of
the data in $X$.

Here we study an ensemble of $M$ realizations of disorder each with $P$
eigenvalues.
For the SVD analysis we construct
a matrix $X$ of size $M \times P$ where $X_{mp}$ is the $p$ level of
the $m$-th realization. After carrying out SVD on $X$, 
the singular values squared $\lambda_k$ are ranked from the largest
to the smallest. This approach has been applied to the
spectrum of disordered systems in several studies
\cite{fossion13,torres17,torres18,berkovits20}.
As is usual in the SVD analysis in these studies 
the first few $\lambda_k$
($k \leq O(1)$) correspond to global features of the spectra.
Larger SV ($\lambda_k$) show a power law behavior $k^{-\alpha}$ with $\alpha=2$
at the Poisson regime and $\alpha=1$ for the Wigner regime.

Here we will examine whether the large  scale  behavior of the
energy can be  gleaned  with the help of SVD. We will use two
different approaches which will eventually lead to similar results.
In the first, we shall use SVD to perform a  global unfolding.
The second will use the scree plot to tease out a power law behavior
for the relevant  energy scale.

\subsection{Global Unfolding}
\label{s3b}

\begin{figure}
  \includegraphics[width=8.5cm,height=!]{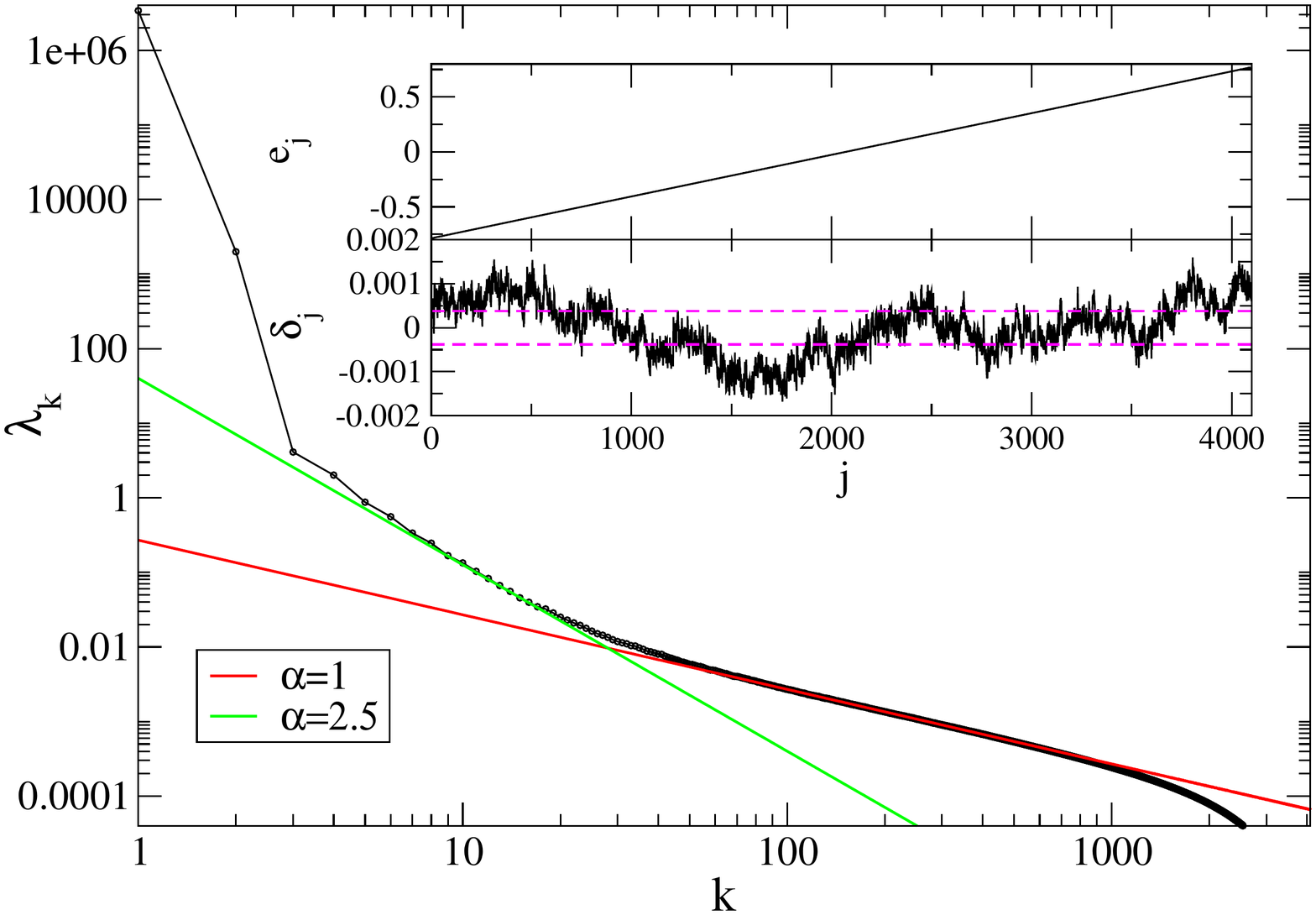}
\caption{\label{fig2}
  A scree plot of the ranked singular values for $M=4096$ different realization
  of the 3D case of size $L=28$, where $P=4096$ eigenvalues around the center
  of the band are considered. The first two modes $k=1,2$ are clearly
  orders of magnitude larger than the rest. Lower modes seem to follow a
  power law $\lambda_k \sim k^{\alpha}$. For $3<k<20$, $\alpha=2.5$ fits well
  while for $k>50$, $\alpha=1$.
  Inset: The contribution of the first couple of modes
  to the $j$-th eigenvalue of a particular realization $e^{i=1}_j$ and
  the contribution of the remaining modes $\delta^{i=1}_j$. The dashed
  magenta line
  indicates a range of $\pm \Delta$ around zero.
}
\end{figure}

The idea behind global unfolding using SVD is to filter out the low modes
which represent the global behavior, while retaining the lower modes that 
encode local fluctuations. We shall illustrate the global unfolding procedure
for an ensemble of realization for the 3D case of size $L=28$.
As previously described, we construct a matrix  $X$, where each row
contains $P=4096$ eigenvalues around the center of the band for each
realization and $M=4096$ columns representing the different
realizations. Matrices $U$,$V$ and $\Sigma$ are numerically extracted
and the $r=M$ diagonal terms $\sigma_k$ are ranked according to amplitude,
from the
largest to the smallest. The matrices $X^{(k)}$ are constructed out of
$U$ and $V$, paying attention to the correct sign \cite{bro08}.
As can be seen in Fig. \ref{fig2}
where the SVD values of $\lambda_k=\sigma_k^2$ are plotted,  
the first couple of modes $\lambda_{k=1,2}$ are clearly orders of magnitude
larger than the lower modes. This is a feature common to  all the cases
considered here. Thus, we may attribute the global features of the
spectrum to the first two modes and  the local fluctuations to the rest.

We define the contribution of the first couple of modes
to the $j$-th eigenvalue of the $i$-th realization as
$e^{i}_j =\sum_{k=1}^2 \sigma_k X_{i,j}^{(k)}$ and the contribution of the rest of
the modes is $\delta^{i}_j =\sum_{k=3}^r \sigma_k X_{i,j}^{(k)}$.
An illustration of the behavior of $e^{i=1}_j$ and $\delta^{i=1}_j$
for the first realization in the ensemble is presented  in the inset of
Fig. \ref{fig2}. It is obvious that  $e^{i=1}_j$ corresponds to
the linear increase of the eigenvalues as function of $j$ expected
in the Anderson model around the center of the band (at zero energy).
Thus, the broad features of the spectra are captured by these two modes.
The local  fluctuations are captured by $\delta^{i=1}_j$,  and one can see the
fast short range fluctuations, but also some longer range ones.
Estimating the mean level spacing $\Delta$ from $e^{i=1}_j$ and comparing it
to $\delta^{i=1}_j$ (see lower inset Fig. \ref{fig2}) further strengthens
the case for longer range fluctuations.

This behavior leaves a very clear mark on the scree plot. Fitting the lower modes
to a power law $\lambda_k \sim k^{\alpha}$, results in two distinct regions.
For $3<k<20$, a power of $\alpha=2.5$ fits well, while $50<k<1000$,
suggests $\alpha=1$. Intuitively, one would guess that modes
$3<k<20$ correspond to longer energy scales for which $E>E_T$ while
$k<50$ to shorter energy scales. Nevertheless, one would like to confirm
this assertion.

\begin{figure}
\includegraphics[width=8.5cm,height=!]{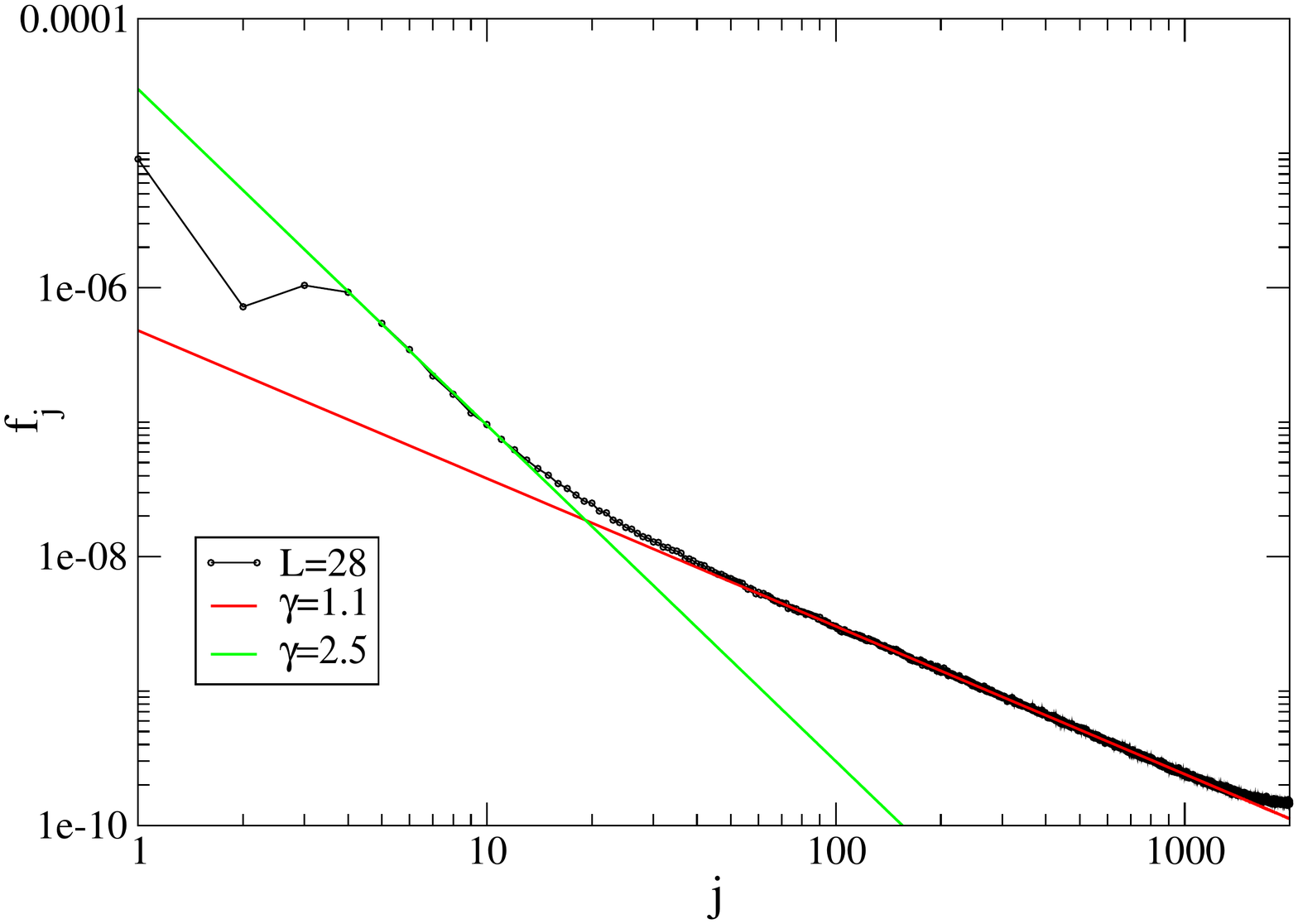}
\caption{\label{fig3}
The power spectrum $F_k$ for the ensemble studied in Fig. \ref{fig2}.
Low frequencies ($4<k<15$, corresponding to large energy scales) follow a
power law $F_k \sim k^{\gamma}$, with $\gamma=2.5$, while
high frequencies ($k>50$, corresponding to small energies), $\gamma=1.1$,
within the expected slope for Wigner-Dyson statistics.
 }
\end{figure}

There has been much work devoted to studying the expression of the
statistics of local fluctuations on the power spectrum of these
fluctuations. It has been shown that the power spectrum of the local
fluctuation of chaotic systems is different than the power spectrum
of integrable systems
\cite{relano02,faleiro04,gomez02,pachon18,gomez05}.
Specifically, the power spectrum of the local fluctuations
for each realization is defined as:
\begin{eqnarray} \label{hamiltonian} 
  F^i_k =  \left| \frac{1}{r} \sum_{j=1}^r \delta^{i}_j
  \exp\left(\frac{-2 \pi i k j}{r}\right)\right|^2, 
\end{eqnarray}
and averaging over all realizations $F_k= \langle F^i_k \rangle$.
For chaotic systems $F_k  \sim k^{-1}$, while for integrable (localized)
systems $F_k  \sim k^{-2}$. The power spectrum of  the local fluctuations
for the 3D case of size $L=28$ is  presented in Fig. \ref{fig3}. As for
the singular value modes, two regimes are apparent. The high frequencies
follow a power law $F_k  \sim k^{-\gamma}$, with $\gamma =1.1$, while
after a crossover a range of low frequencies fit to $\gamma =2.5$.
Thus the high frequencies ($k>50$, corresponding to small energies), the
behavior of the power spectrum is close to what was observed
in other chaotic (GOE) systems 
\cite{relano02,faleiro04,gomez02,pachon18,gomez05}. There are
a couple of interesting observations that one can draw from the behavior
of the low frequencies. The first has to do with the equivalence
between the power laws of the SVD scree plot for low modes and for the 
for the power law at low frequencies, i.e., $\alpha=\gamma=2.5$, and
for the high modes and frequencies $\alpha \sim \gamma \sim 1$.
Such correspondence between the power law of  the SVD modes and power spectrum
frequencies has been noted for the energy spectrum in Refs.
\onlinecite{fossion13,torres17,torres18,berkovits20},
and elucidated in Ref. \onlinecite{bozzo10}.
The correspondence also  determines the energy scale  of the singular values.
The $k$-th Fourier transform frequency corresponds to an energy scale
$P \Delta/2k$, thus the region for which GOE statistics holds is of
order  of  $40 \Delta$, not to far from the estimation of the Thouless
energy obtained via the number variance. As can be seen in Fig. \ref{fig2}, also
the singular value modes follow the GOE expectation ($\alpha=1$) up to
$k=50$. Moreover, both curves show a similar behavior and one may assume that
the scree plot depicts the same  physics as the power spectrum of the
globally unfolded  energy spectrum and that the energy scales probed by
the modes of the SVD are similar to the energy scales of the Fourier transform.

\begin{figure}
\includegraphics[width=8.5cm,height=!]{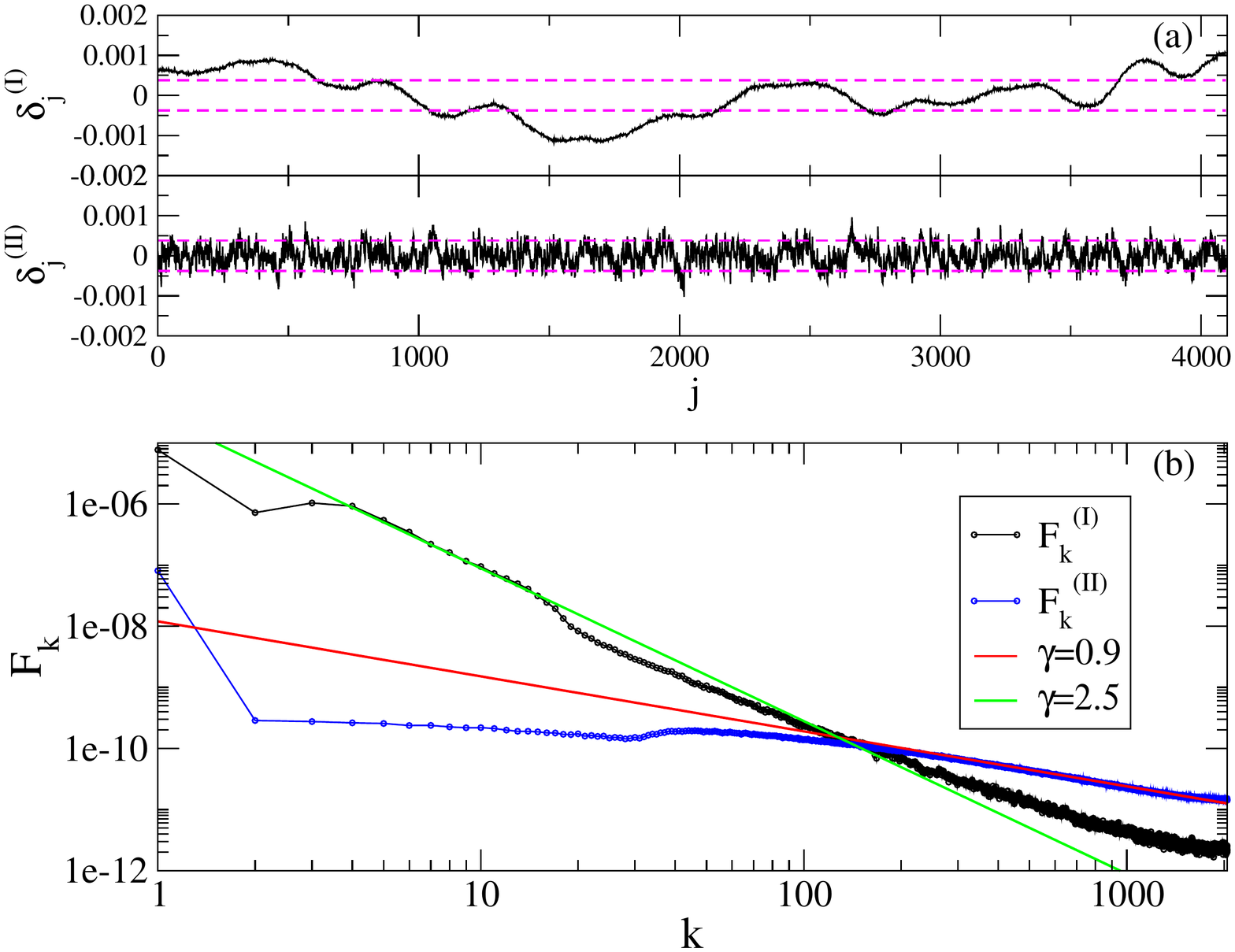}
\caption{\label{fig3a}
(a) The contribution of the lower modes ($3<k<30$) $\delta^{i=1 (I)}_j$
and the higher modes ($31<k<r$) $\delta^{i=1 (II)}_j$
to the $j$-th eigenvalue of a particular realization $i=1$. The dashed
magenta line indicates a $\pm \Delta$ (averaged level spacing).
A clear difference between the low modes which encode long range fluctuations
and  the higher modes  which represent short range fluctuations is apparent.
(b) The power spectrum $F_k^{(I,II)}$ of $\delta^{i (I,II)}_j$.
The low modes power spectrum $F_k^{(I)}$ show a slope of $\gamma=2.5$ for
the low frequencies ($4<k<20$) and tappers of for higher frequencies.
The power spectrum  $F_k^{(II)}$ for the higher modes reveals that these
modes correspond to
high frequencies at the range $k>100$, with a slope $\gamma=0.9$.
 }
\end{figure}

In order to substantiate the proposed connection between the energy scale
and the SVD mode number,
we split the
the contribution the modes into two parts:
$\delta^{i (I)}_j =\sum_{k=3}^{30} \sigma_k X_{i,j}^{(k)}$ and
$\delta^{i (II)}_j =\sum_{k=31}^r \sigma_k X_{i,j}^{(k)}$.
As can be seen in Fig. \ref{fig3a}a for the same realization
presented in the inset of Fig. \ref{fig2}, $\delta^{i (I)}_j$ indeed depicts
longer range fluctuations, while $\delta^{i (II)}_j$ portrays short scale
fluctuations. This could  be confirmed  by the power spectrum  of
the fluctuations $\delta^{i (I,II)}_j$. For the low singular value modes the
corresponding power spectrum (see Fig. \ref{fig3a}b)
frequencies are in the range of $4<k<20$ with the same
power law $\alpha=\gamma=2.5$, while for  the higher modes the corresponding
frequencies  are at $k>100$ with a slope $\gamma=0.9$.
Thus one can reasonable conclude that
low modes in the SVD probe the large energy scales of the spectrum.

Another observation is that both for the power spectrum and
for the SVD scree plot the lower frequencies/modes exhibit a power law
with a slope of $1+d/2$. The slope of the power spectrum
could be generally associated to the value of the power of the variance
\cite {mcdowell07,krapf18}. As detailed in the appendix, indeed,
the expected value of the power spectrum  $\gamma=1+d/2$ can be
analytically explained.
In the next sub-section we will further substantiate these observation.

\subsection{Scree Plot}
\label{s3c}

As we have previously seen the scree plot of the singular values
characterize the behavior of the large energy scale by showing a  power law
behavior of the low  modes corresponding to $1+d/2$ power,  clearly
distinct from $\alpha=1$ seen for higher modes. Here we would like to
check  whether this behavior is robust for different system sizes, disorder
strength,  ranges of the spectrum and dimentionality.

\begin{figure}
\includegraphics[width=8.5cm,height=!]{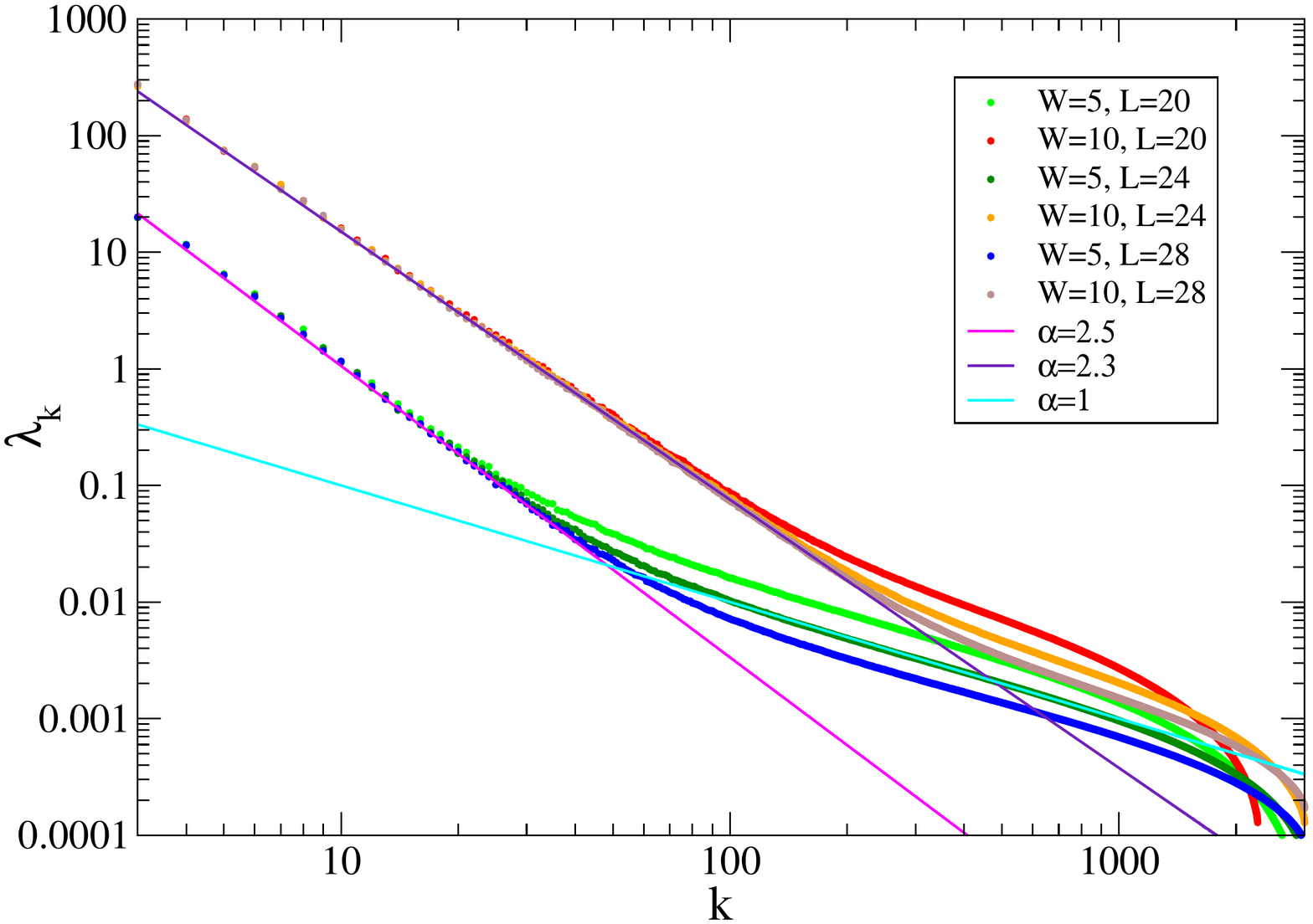}
\caption{\label{fig4}
  The SVD modes scree plot for 3D systems  where an ensemble of $M=3000$
  realizations of disorder and sizes $L=20, 24, 28$ are considered. In all
  cases $P=L^3/2$ eigenvalues around the center of the band are taken into
  account. Two  different  strength of disorder $W=5$ and $W=10$ for all sizes
  are presented. Lines depict different slopes $\lambda_k=k^{-\alpha}$, with
  $\alpha=2.5,2.3,1$.
 }
\end{figure}

First we continue to present results for the SVD modes scree plot
for 3D samples in Fig. \ref{fig4}. All results are for an ensemble
of $M=3000$ different realizations at each size and disorder strength.
Three different  sizes $L=20,24,28$ are considered, each for
two different strength of disorder $W=5$ (as we saw
corresponds to $n(E_T)=g=20$) and $W=10$ ($n(E_T)=g=5$,  makes sense since
$E_T \sim 1/W^2$). Thus, for the $W=5$ samples we are deep in the metallic
regime where the Altshuler and Shklovskii's predictions are expected to hold,
while for $W=10$ we are already closer to the localized regime ($g=1$).
Indeed, one can see  that the range of modes for
which the GOE behavior holds($\alpha \sim 1$) is much lager for the
weaker disorder. The weak disorder singular values fall on top of each other
for the lower modes, with a slope of $\alpha=2.5=1+d/2$. Then for higher modes
the slope switches to the GOE behavior ($\alpha \sim 1$) where the value
of $k$ for which the switch occurs is higher as the system size increases.
We speculate that this is the result of  the fact that for larger systems
there are more eigenvalues in the range of $L^3/2$. We shall further
substantiate this assertion shortly. A similar behavior is  seen for
the stronger disorder ($W=10$) although the slope deviates a bit
from  $\alpha=1+d/2$ and is closer to $\alpha=2.3$. This is not  surprising
since the prediction in  Ref. \onlinecite{altshuler86} were obtained using
diagrammatic reasoning, strictly valid only deep in the metallic regime
($g \gg 1$).

\begin{figure}
\includegraphics[width=8.5cm,height=!]{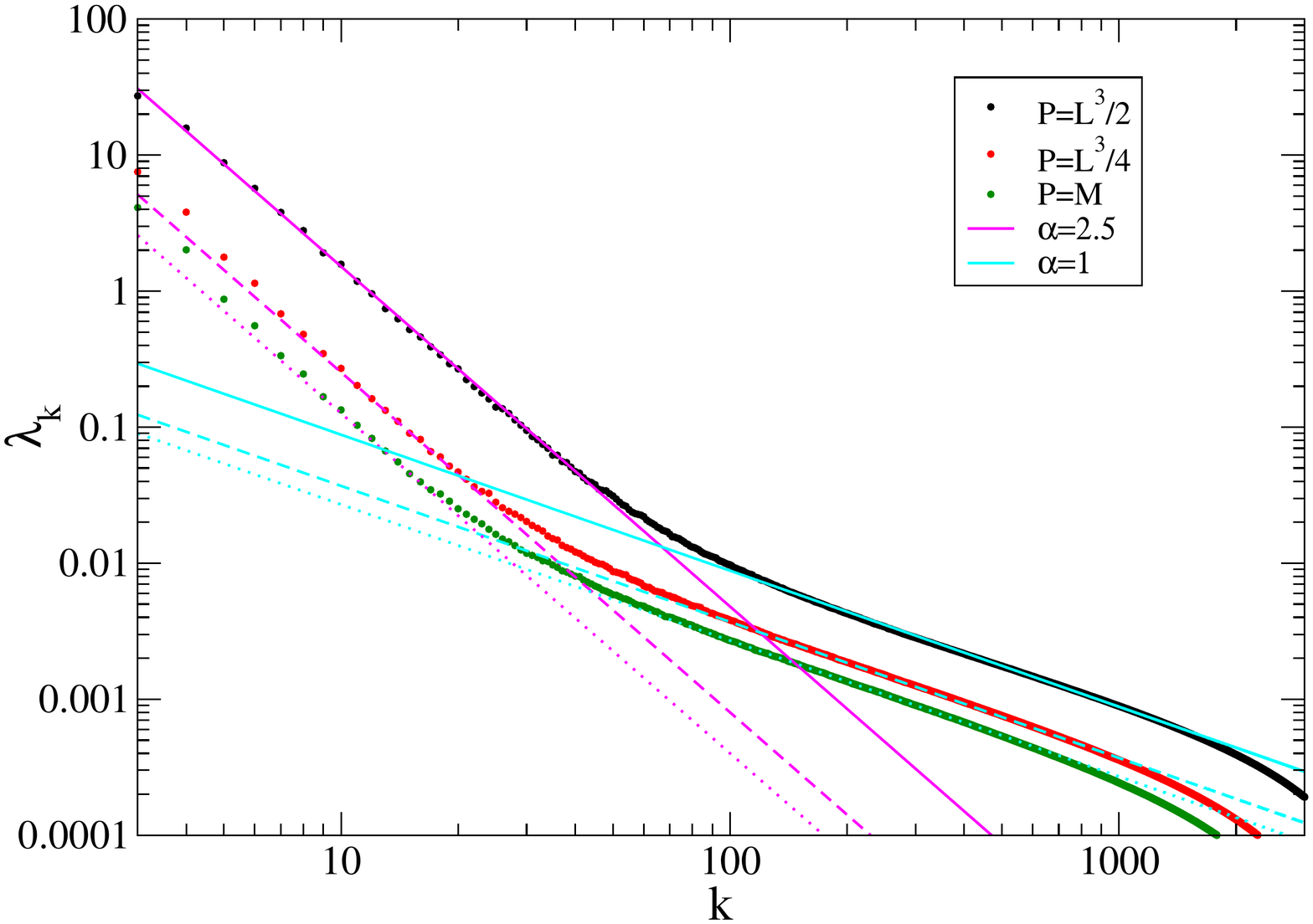}
\caption{\label{fig5}
  The SVD modes scree plot for 3D systems ($M=4096$
  realizations) of size $L=28$.
  Three different ranges of
  eigenvalues straddling the center of the band are presented:
  $P=L^3/2=10976$, $P=L^3/4=5488$, and $P=M$.
  account. The qualitative behavior of the slopes, i.e., $\alpha=2.5$
  (magenta line) for
  low modes and $\alpha=1$ for higher modes (cyan line),
  does not change,  although
  the crossover occurs at smaller values of $k$ as $P$ becomes smaller.
 }
\end{figure}

In Fig. \ref{fig5} we examine the influence of the change in the range
of the eigenvalues, $P$ on $\lambda_k$. Indeed, the main influence of
narrowing the range of $P$ is to shift the crossover from  the $\alpha=2.5$
slope to  the GOE  $\alpha=1$ slope to lower values of $k$. This makes sense,
since the smaller the range, the smaller is the number of energies larger than
the Thouless energy in this range. Thus, when one wants to focus on energies
beyond the Thouless energies, and there is a limit on the ensemble size
$M$, one should  expand
$P$ to the largest available range,  even if $P \gg M$.

\begin{figure}
\includegraphics[width=8.5cm,height=!]{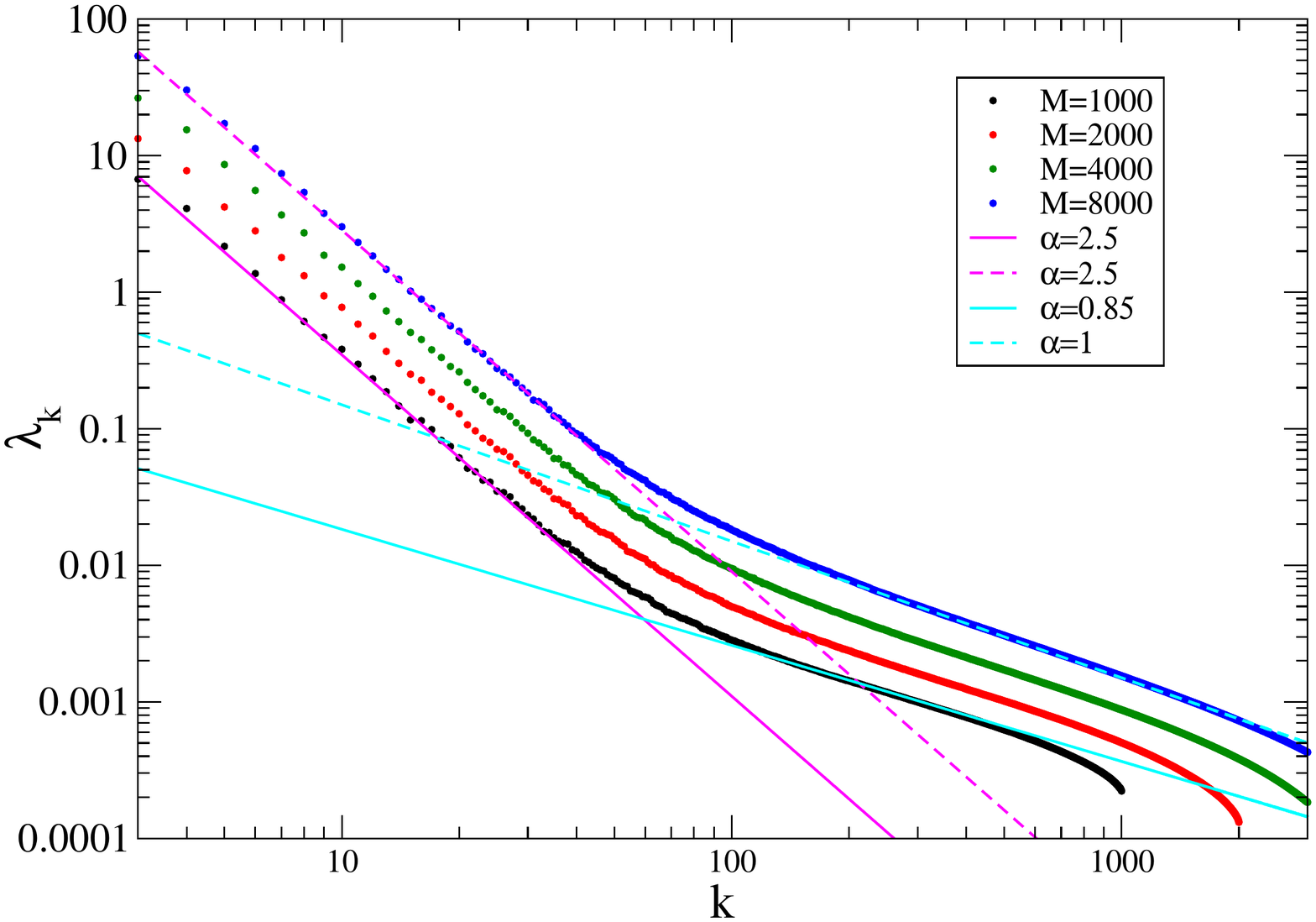}
\caption{\label{fig5a}
  The SVD modes scree plot for 3D systems of size $L=28$  and a different
  number of realizations $M=1000,2000,4000,8000$ for the same
  $P=L^3/2=10976$.
  The behavior of the low  mode slopes is not sensitive to
  to the ensemble size and a slope of $\alpha=2.5$ is seen for all values
  of $M$.
  Lower modes are more suscept to the number of realizations. For small values
  of   $M=1000$ the slope fits
  $\alpha=0.85$ for $M=1000$ at higher modes, while it shifts to
  $\alpha=1$ at $M=8000$.
 }
\end{figure}

The number of the realizations taken in the ensemble, $M$, also plays a role
in the behavior of the SVD modes. As can be seen in Fig. \ref{fig5a}
the slope for the low mode
does not depend on the number of realizations in the ensemble $M$ and remains
$\alpha=2.5$ for all values of $M$. On the other hand, for the high modes the
slope varies from $\alpha=0.85$ at $M=1000$ to  $\alpha=1$  for the largest
number of realizations $M=8000$, in line with the predictions
for the Wigner-Dyson statistics. Such behavior  
has been previously seen  in the study of the 
generalized Rosenzweig-Porter where the scree plot  of the SVD modes for
large $k$ (the GOE regime) also follow a slope of $\alpha \sim  0.8$
\cite{berkovits20}  for small  values of $M$. This behavior was attributed
there to the fact that there $M << P$. This fits well with our current
results where  as $M$ grows $\alpha$ is closer to one.

\begin{figure}
\includegraphics[width=8.5cm,height=!]{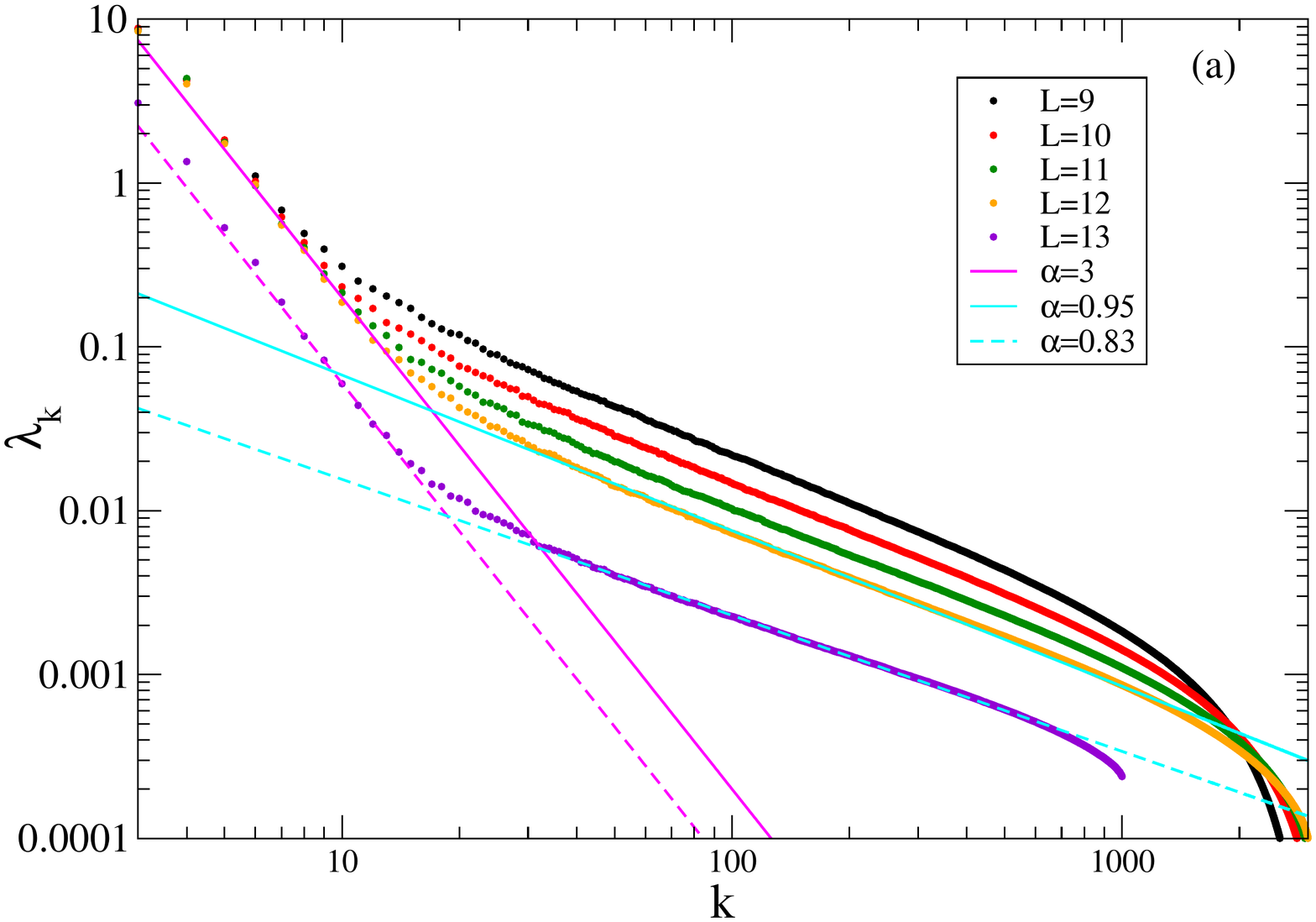}
\includegraphics[width=8.5cm,height=!]{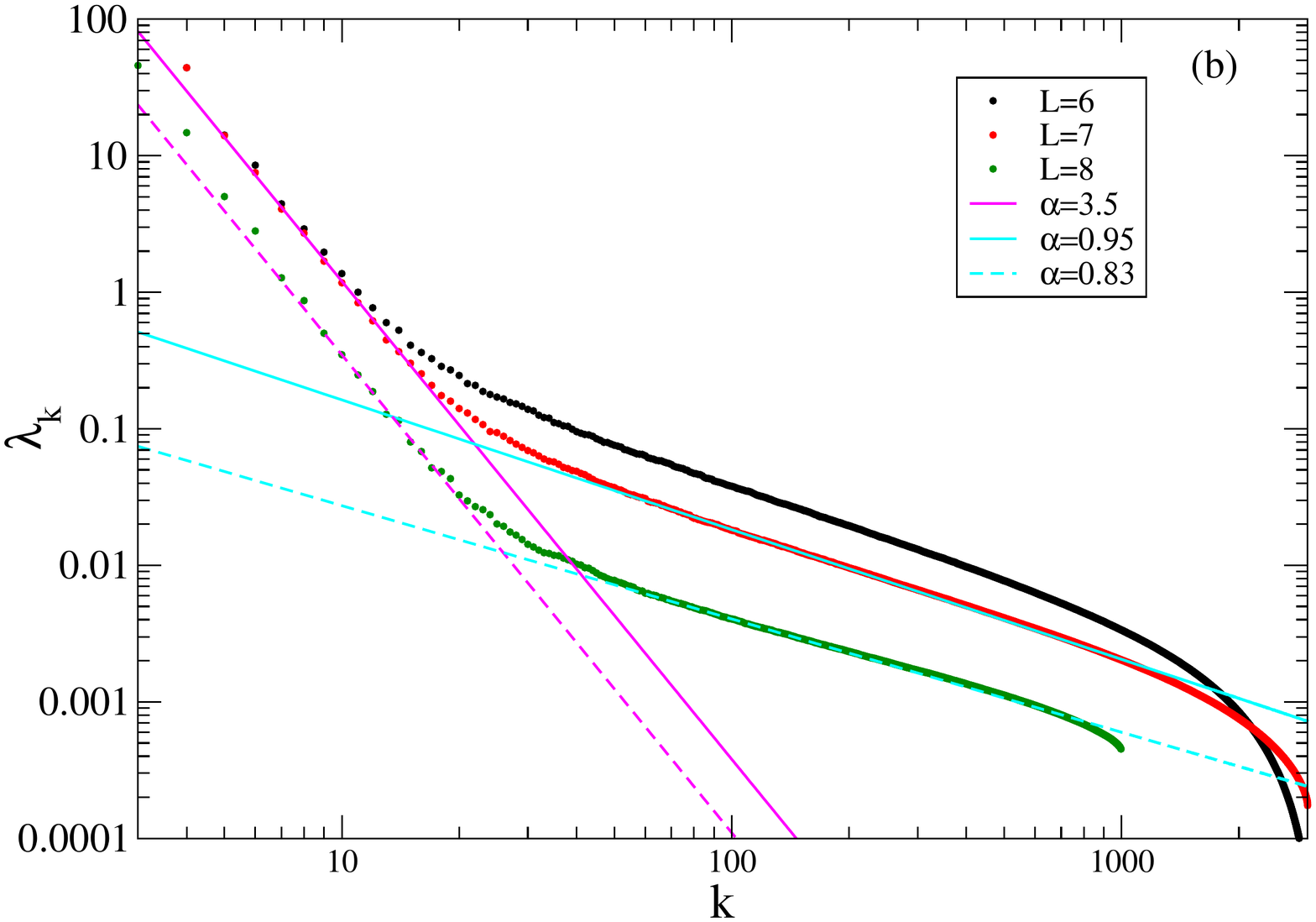}

\caption{\label{fig6}
  The SVD modes scree plot for 4D (a) and 5D (b)
  systems ($M=3000$
  realizations except for the largest size in each case where $M=1000$)
  of sizes $L=9, 10, 11, 12, 13$ for the 4D case
  and $L=6, 7, 8$ for 5D realizations
  of disorder $W=5$ (4D) and $W=10$ (5D). The number of eigenvalues $P=L^3/2$.
  For both dimesnsionalities the low modes slopes follow
  a $1+d/2$ behavior, i.e., $\alpha=3$ for $d=4$ and $\alpha=3.5$ for $d=5$,
  and a slope close to one ($\alpha=0.95$) for the higher modes except at the
  largest  sizes where $\alpha=0.83$. 
 }
\end{figure}

Finally, we wish to examine the dependence on dimensionality of the  SVD modes.
As we have seen for $d=3$ and argued analytically, we expect to observe a slope
of $\alpha=1+d/2$ for the lower modes crossing over to a slope of
$\alpha =1$ at higher modes. Indeed, the scree plot shown in Fig \ref{fig6}
confirms that the slope
of the low modes corresponds to $\alpha=3$, and $\alpha=3.5$
for $d=4$ and $d=5$, as expected from the number variance 
behavior at large energies predicted in Ref. \onlinecite{altshuler86}.
Higher modes show a slope of $\alpha=0.95$ for both $d=4$ and $d=5$,
close to the expected value of $\alpha=1$, except for the largest length at each
dimensionality for which the number of realizations $M=1000$ is smaller than
for the other length,  and the slope is $\alpha=0.83$. This is in
line with the behavior shown in Fig. \ref{fig5a}.

\section{Discussion}
\label{s4}

The detection of the Thouless energy in the spectrum of weakly
disordered chaotic systems has long been achieved by detecting the
deviation  from the expected Wigner Dyson logarithmic dependence of the
number variance. Thus, it could be assumed that the number variance
will also reveal the behavior of the spectrum at energy scales beyond the
Thouless energy derived by Altshuler and Shklovskii \cite{altshuler86}.
Indeed a  stronger than linear dependence has been frequently  observed,
nevertheless, extracting the expected power law behavior from  the
number variance after local unfolding has turned out to  be far from
trivial. As has been shown here, the number variance does indeed
show a  $\langle n^2(E) \rangle \sim \langle n(E) \rangle^{\beta}$
behavior  for a significant range of levels.  Although the value
of $\beta$ rises  as the size of sample increases towards the expected
$d/2$  value, it remains hard to extrapolate a value with the largest samples
we are able to compute.

By taking the route of the SVD, it is possible to overcome
these difficulties. The SVD essentially decompose the spectrum
to modes, where the low  modes  (large amplitudes) capture the longer
range features. As we have shown, the SV modes are in a sense similar to the
Fourier transform frequencies, and show similar regularities of the
frequencies and modes. Nevertheless, the SVD saves the need  to first unfold
and then perform a power analysis over all realizations and finally
average, thus it is a much more concise method. Moreover, since
the contribution of the lowest modes ($e^{i}_j $) filtered out
is custom set for each realization $i$, one overcomes the problem  of
individual realization global variations raised in Ref. \onlinecite{sierant19}.
Thus, either by unfolding with
SVD  and then performing a power spectrum, or by directly examining the
singular values using the scree plot, it is possible to extract the properties
of the energy spectra beyond the Thouless energy and to see the predictions
of Ref. \onlinecite{altshuler86} clearly  hold.

This success might encourage the  application  of the SVD method to other 
systems for  which interesting long range properties of the energy spectrum
are expected  such as the SYK model and systems which show
many body localization.

\appendix*
\section{Connection between the number  variance and the power spectrum}

Here we aim to show that the relation between the slope of
the power spectrum $\langle F_k \rangle \propto k^{-\gamma}$ and
the slope of the number variance 
$\langle \delta^2 n(E) \rangle \propto \langle n(E) \rangle^{d/2}$ is
$\gamma = 1+d/2$.

Following McDowel {\it et. al.} \cite{mcdowell07} the local fluctuations
$\tilde \delta^{i}_j = \delta^{i}_j/\Delta$ may be rewritten as:
\begin{eqnarray} \label{cos} 
    \tilde \delta^{i}_j = \frac{1}{k_{max}-k_{min}} \sum_{k=k_{min}}^{k_{max}}
    \sqrt {2 F^i_k} \cos(\mathcal{K} j + \phi^i_k),
\end{eqnarray}
  where $k_{min},k_{max}$ is the range for which the power spectrum
  exhibits a particular power law behavior with slope $\gamma$,
  $\mathcal{K}=2 \pi k/r$, and $\phi_i$ is a  phase. The number variance
  $\delta^2 n_j^i(E)$ for a  particular realization $i$
  where the energy window $E$ starts at the energy of the averaged $j$th
  eigenvalue $j \Delta$ and ends at $(j+l) \Delta$ ($E= l\Delta$)
  could be written as:
\begin{eqnarray} \label{nv} 
  \tilde \delta^2 n_j^i(l \Delta)=  (\tilde \delta^{i}_{j+l} -
  \tilde \delta^{i}_j)^2.
\end{eqnarray}
Substituting $\tilde \delta^{i}_j$ by
Eq. (\ref{cos}) and averaging over the beginning
of the energy window $j$ and the phase $\phi_k$ one obtains:
\begin{eqnarray} \label{nv1} 
  &\langle \delta^2 n(l \Delta) \rangle = \frac{1}{r-l}\sum_{j=1}^{r-l}
  \frac{1}{(k_{max}-k_{min})^2} \\ \nonumber
  &\sum_{k=k_{min}}^{k_{max}} \sum_{k \prime=k_{min}}^{k_{max}}  
  \frac{2 F_k}{(2\pi)^2}\int_0^{2\pi} d \phi_k d \phi_k\prime \\ \nonumber
  &\big(\cos(\mathcal{K} (j+l) + \phi_k)-\cos(\mathcal{K} j + \phi_k)\big)
  \\ \nonumber
  &\big(\cos(\mathcal{K} (j+l) + \phi_k\prime)-\cos(\mathcal{K} j + \phi_k\prime)\big),
\end{eqnarray}
resulting in:
\begin{eqnarray} \label{nv2} 
  &\langle \delta^2 n(l \Delta) \rangle = \frac{1}{r-l}\sum_{j=1}^{r-l}
  \frac{1}{k_{max}-k_{min}} \sum_{k=k_{min}}^{k_{max}}\\ \nonumber
  &\frac{2 F_k}{2\pi}\int_0^{2\pi} d \phi_k \big(
  \cos^2(\mathcal{K} (j+l)+ \phi_k)+\cos^2(\mathcal{K} j + \phi_k)\\ \nonumber
  &-2  \cos(\mathcal{K} (j+l) + \phi_k)\cos(\mathcal{K} j + \phi_k) \big).
\end{eqnarray}
Performing  the integration over $\phi_k$ and summation over $j$
retaining only the $l$ dependent part one obtains:
\begin{eqnarray} \label{nv3} 
  \langle \delta^2 n(l \Delta) \rangle \sim 
  \frac{1}{k_{max}-k_{min}} \sum_{k=k_{min}}^{k_{max}} 
  F_k \cos(\mathcal{K}l).
\end{eqnarray}
Replacing the summation with an integration and using the power law
dependence of the power spectrum leads to:
\begin{eqnarray} \label{nv4} 
  \langle \delta^2 n(l \Delta) \rangle \sim
  \int_{{\mathcal{K}}_{min}}^{{\mathcal{K}}_{max}} d \mathcal{K}
  \mathcal{K}^{-\gamma} \cos(\mathcal{K}l) \sim l^{\gamma-1}.
\end{eqnarray}
Thus since  followin Altshuler and Shklovskii \cite{altshuler86} 
$\langle \delta^2 n(l \Delta) \rangle \sim l^{d/2}$, one concludes that the
power spectrum should exhibit a slope $\gamma=d/2+1$.

\end{document}